\documentclass[epj]{svjour}
\usepackage{graphicx}
\begin{document}
\title{Conductance of tubular nanowires with disorder}
\author{Lloren\c{c} Serra\thanks{On sabbatical leave from 
Institut de F\'{\i}sica Interdisciplinar i de Sistemes Complexos IFISC (CSIC-UIB)
and Departament de F\'{\i}sica, 
Universitat de les Illes Balears, E-07122 Palma de Mallorca, Spain.} 
\and Mahn-Soo Choi
%
}                     
\institute{Department of Physics, Korea University, Seoul 136-701, Korea}

\date{20 July 2009}
%
\abstract{ We calculate the conductance of tubular-shaped
nanowires having many potential scatterers at random positions.
Our approach is based on the scattering matrix formalism
and our results analyzed
within the scaling theory of disordered conductors. 
When increasing the
energy the conductance
for a big enough number of impurities
in the tube
manifests a systematic evolution from 
the localized to the metallic regimes. Nevertheless, a conspicuous drop in conductance 
is predicted whenever a 
new transverse channel is open. 
Comparison with the semiclassical calculation leading 
to purely ohmic behavior is made.
} 
\maketitle
\section{Introduction}

Electronic transport in nanostructures is often 
affected by the presence of impurities. 
Understanding impurity induced disorder has been a major concern
in the field of quantum transport for many years. 
A vast literature on this problem has accumulated since Anderson's
discovery of the phenomenon of weak localization \cite{Anderson}
and the proposal of the scaling theory \cite{scaling}. A lengthy
introduction to the field is out of the scope of this work but the 
reader is addressed to Refs.\ \cite{rev1,rev2,rev3,rev4} for reviews on the
topic.

In this paper we address a specific nanostructure geometry, namely
a two-dimensional (2D) electron gas on the surface of a tube. Our motivation to 
study this particular nanostructure shape is mainly due to the similarity
with rolled-up semiconductor quantum wells recently 
fabricated \cite{expertube,et2,et22,et3,et4} 
and studied \cite{theotube,tt2,tt3}.
Here, it is also worth mentioning the apparent resemblance with electrons
in carbon nanotubes, although these latter systems are generally of 
smaller dimensions.
We also stress that this geometry is
particularly adapted to the theoretical modeling of random impurities.
Indeed, by means of longitudinal translations and rotations, i.e., the tube 
symmetries, one can always relate the position of two impurities on the 
tube surface.

We have implemented the numerical modeling of the disordered
nanotubes using the scattering matrix formalism. 
In this approach, the solution of the Schr\"odinger equation is obtained by
treating each scatterer as a black box characterized by its  
transmission and reflection amplitudes. The solution to the many impurity
problem is then found by adequately composing the single impurity scattering
matrices. Our approach is similar to that of Cahay, McLennan and Datta
for planar wires \cite{cahay}. A practical difference, however, 
is that we do not present the problem as a matrix recursion but,
instead, as a global sparse linear system.

When increasing the number of propagating transverse modes the system 
behavior evolves from purely one dimensional (1D) in the one-mode limit towards 
2D in the infinite-mode limit. In this paper we focus on the 
few-mode regime, or quasi-1D limit, monitoring the evolution of the conductance 
for energies allowing up to five propagating modes. In all cases the disordered 
wire is characterized by the asymptotic exponential localization of the wave 
function. However, it will be shown below that depending on the energy remarkable
variations of the localization properties are predicted.

The scaling theory of disordered wires implies that, given a
wire length $L$, wire conductance is a random variable $\tilde{G}$, characterized
by a certain probability distribution $P_{L,\gamma}(\tilde{G})$.
Randomness appears due to the infinite ways in which 
the number of impurities ${\cal N}$ can be arranged (each particular 
arrangement is usually referred as a disorder realization).
Besides $L$, the probability distribution is also characterized
by parameter $\gamma$, usually known as Lyapunov exponent \cite{rev2}.  The relation 
between the Lyapunov exponent and the mean conductance strongly depends on 
the wire length. In other words, the form of $P_{L,\gamma}(\tilde{G})$
changes strongly with $L$. The precise definition of $\gamma$ involves 
averaging a function of the random conductance over all disorder realizations   
in the limit of a very long wire; namely
\begin{equation}
\gamma = \lim_{L\to\infty}{\langle\tilde\gamma\rangle}\; ,
\end{equation}
where we have defined 
\begin{equation}
\label{gamt}
\tilde\gamma = \frac{1}{L}\log\left(1+\frac{G_0}{\tilde{G}}-\frac{1}{M}\right)\: .
\end{equation}
In Eq.\ (\ref{gamt}), $G_0\equiv 2e^2/h$ represents the conductance quantum and
$M$ is the number of propagating modes in the wire. 
For a given $L$ and disorder realization the system is characterized by a single
value of $\tilde G$ and its associated $\tilde\gamma$. Changing the disorder
realization these variables acquire the statistical meaning implied above.
Clearly, $\tilde\gamma$ is then a random variable whose mean, in the long wire limit,
gives the Lyapunov exponent.

The inverse of the Lyapunov exponent gives the localization length $\ell=\gamma^{-1}$. 
This length is physically relevant since it specifies how $\tilde{G}$ and $\tilde\gamma$
are distributed for a given $L$. Two limiting regimes are known.
\begin{itemize}
\item[a)] $L\gg \ell$ ({\em localized} regime): $\tilde\gamma$ is normally (Gaussian) 
distributed with mean value $\gamma$ and width $\sigma^2_{\rm loc}=2\gamma/L$. In this limit we then have
\begin{equation}
\label{pd1}
P_{L,\gamma}(\tilde{G})=\frac{1}{\sqrt{2\pi}\sigma_{\rm loc}}
e^{-(\tilde\gamma-\gamma)^2/2\sigma^2_{\rm loc}}
\frac{d\tilde\gamma}{d\tilde{G}}\;,
\end{equation}
where $\tilde\gamma$ as well as $d\tilde\gamma/d\tilde{G}$ can be obtained from 
$\tilde{G}$ using Eq.\ (\ref{gamt}). 

\item[b)] $L\ll\ell$ ({\em metallic} regime): $\tilde{G}$ is normally distributed 
\begin{equation}
\label{pd2}
P_{L,\gamma}(\tilde{G})=\frac{1}{\sqrt{2\pi}\sigma_{\rm met}}e^{-(\tilde{G}-G)^2/2\sigma^2_{\rm met}}
\;,
\end{equation}
with mean value given by 
\begin{equation}
\label{meang}
G=\frac{G_0}{e^{\gamma_L L}-1+\frac{1}{M}}\; ,
\end{equation}
and a constant variance. Indeed, universality in the value of
$\sigma_{\rm met}$ is the reason why this limit is also dubbed 
the universal conductance fluctuation regime. 
In our case in which time reversal and spin rotation symmetries 
are fulfilled this value is $\sigma_{\rm met}=\sqrt{\frac{2}{15}}G_0$ \cite{rev1}.
Notice that in Eq.\ (\ref{pd2}) we have considered the possibility that the mean
value $\gamma_L\equiv\langle\tilde\gamma\rangle$ may not be fully
converged yet to the Lyapunov exponent for a given $L\ll\ell$. 
\end{itemize}

Purely Ohmic behavior is characterized by a strict linearity of the resistance with the
wire length. Such a behavior is obtained in a semiclassical description, where 
any possibility of
quantum interference is neglected, and the scatterers are composed incoherently.
Total transmission in this case does not depend on the impurity
positions \cite{notesc}, but only on its number ${\cal N}$ which is
fixed for a certain length $L$.
The semiclassical conductance $G_{\rm sc}(L)$ is then fully deterministic and can be written in terms 
of a single parameter $\alpha$ as
\begin{equation}
\label{Gsc}
G_{\rm sc}(L)=\frac{G_0}{\alpha L + \frac{1}{M}}\; .
\end{equation}
Deep in the metallic regime the mean quantum conductance also has an 
Ohmic behavior, as seen immediately from
Eq.\ (\ref{meang}) assuming $\gamma_L=\gamma$
and expanding $e^{\gamma L}\approx 1+\gamma L$.
Notice, however, that this Ohmic dependence may be
different from the semiclassical one since we may have 
$\alpha\ne\gamma$, as will be shown below for particular cases. 
Including the next-order contribution from the exponential and assuming many 
propagating modes such that $\frac{1}{M}\ll\gamma L\approx \alpha L \ll 1$ we
find
\begin{equation}
\label{diff}
\langle \tilde{G}\rangle-G_{\rm sc}(L) =G_0
\frac{(\alpha-\gamma)L-\frac{1}{2}(\gamma L)^2}{(\alpha L)(\gamma L)}\;.
\end{equation}
If $\alpha=\gamma$, Eq.\ (\ref{diff}) predicts 
$\langle \tilde{G} \rangle=G_{\rm sc}(L)-G_0/2$, but this need not be true 
in general.

In this paper we calculate the localization length of disordered tubes
and its evolution with energy covering from one to five propagating modes. 
The above prediction of scaling theory in the 
localized and metallic regimes is shown to be fulfilled and the
deviation from it in the intermediate regime is explicitly shown.
Given a number of propagating modes we find a steady increase
of localization length with energy. Quite remarkably, however,
a large drop is obtained each time a new mode becomes propagating.
As pointed by Thouless \cite{thouless} 
the localization length is related to the density of states and, therefore, 
we may expect some type of discontinuity in our quasi 1D system. 
This variation in localization length is the cause of similar drops
in the mean conductance of a disordered tube of length $L$.  In this 
case, increasing the energy the tube evolves from a more localized 
towards a more metallic regime. 

The paper is organized as follows. In Sec.\ 2 we present the model and 
our approach to the many impurity scattering problem (in appendix we also 
discuss the solution 
of the single impurity problem).
Section 3 contains 
the results and its discussion while Sec.\ 4 draws the conclusions
of the work. 

\begin{figure}[t]
\centerline{\includegraphics[width=6.5cm,clip]{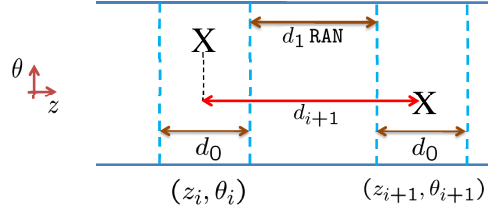}}
\caption{Parameters of the impurity distribution.}
\label{fig1_p}
\end{figure}

\section{Model}
\label{model}

We assume the electrons move on the surface of a cylinder of radius $\rho$.
Disorder is represented by repulsive potential barriers mimicking 
the effect of impurities on the tube surface. The potential is 
assumed of a short range (Gaussian) type
\begin{equation}
\label{vimpu}
V^ {(i)}({\bf r})=V_0 e^{-|{\bf r}-{\bf r}_i|^2/\sigma^2}\; ,
\end{equation}
where ${\bf r}_i$ is a specific impurity position.  
${\cal N}$ such impurities, a rather large number,
are present with positions
$\left\{ {\bf r}_i, i=1,\dots,{\cal N}\right\}$.
All impurities are assumed to be identical, although we would not expect
qualitative differences in the results presented below if some variation 
in $V_0$ and $\sigma$ was considered.

\subsection{Impurity distribution}

Impurities are assumed to be well separated compared with the potential range $\sigma$.
A sequential random 
distribution is then represented by the two parameters $d_0$ and $d_1$ (see a 
sketch in Fig.\ 1). The first impurity is arbitrarily positioned
at $(z_1,\theta_1)=(0,0)$ and successive ones at
\begin{equation}
\left\{
\begin{array}{rcl}
z_{i}&=& z_{i-1}+d_0+d_1\, \verb+RAN+ \\
\theta_{i} &=& 2\pi\, \verb+RAN+ 
\end{array}
\right.
\quad
(i=2,\dots,{\cal N})
\; ,
\end{equation}
where \verb+RAN+ is a standard random number uniformly distributed between 0 and 1.
We define $d_i=z_i-z_{i-1}$, the longitudinal separation between impurities $i$ 
and $i-1$. Obviously, on average $\langle d_i\rangle=d_0+0.5 d_1$ and the total length
$L$ of the disordered tube is accurately given by $L=(d_0+0.5d_1){\cal N}$.  
The idea behind this model is that all physical effects on electronic motion 
by an individual scatterer
have vanished within a distance $d_0$ around the scatterer's position, the 
electron propagating then freely a distance $d_1\verb+RAN+$ to the proximity 
of the next scatterer.

\begin{figure}[t]
\centerline{\includegraphics[width=7.2cm,clip]{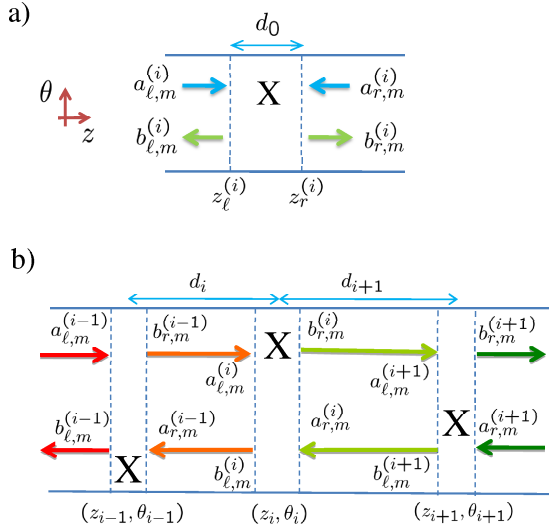}}
\caption{Sketch of input and output amplitudes of the scattering matrix approach for a single
impurity (a) and for many impurities (b). In the latter case input and output amplitudes
of successive impurities are related.}
\label{fig2_p}
\end{figure}

\subsection{Scattering matrix approach}
\label{scatsec}

Let us consider a single impurity, the $i$-th one, say, and write the wave function far from it 
(outside the region of length $d_0$, see Fig. 2a) as
\begin{eqnarray}
\label{wf}
\psi(z,\theta) &=&
\sum_{m=1}^{M}{a_{c,m}^{(i)} \phi_m(\theta) e^{i s_c k_m (z-z_c^{(i)})}} \nonumber\\
&+&
\sum_{m=1}^{M}{b_{c,m}^{(i)} \phi_m(\theta) e^{-i s_c k_m (z-z_c^{(i)})}}\; ,
\end{eqnarray}
where  $M$ is the number of propagating transverse 
modes $\{\phi_m(\theta)\}$.  
In Eq.\ (\ref{wf}) we have introduced a {\em contact} 
label $c=\ell,r$, for left
and right asymptotic regions with boundaries at
$z_\ell^{(i)}$ and $z_r^{(i)}$, respectively. We have also defined $s_\ell=1$
and $s_r=-1$, and the mode wavenumber $k_m$ (see the appendix). 
As it is normally defined \cite{datta},
the scattering matrix relates output current amplitudes to input ones 
\begin{equation}
\label{scatmat}
\left(
\begin{array}{c}
b_{\ell,m}^{(i)}\sqrt{v_m}\\
b_{r,m}^{(i)}\sqrt{v_m}
\end{array}
\right)
=
\left(
\begin{array}{cc}
r_{mn}^{(i)} & {t'}_{mn}^{(i)} \\
t_{mn}^{(i)} & {r'}_{mn}^{(i)} \\
\end{array}
\right)
\left(
\begin{array}{c}
a_{\ell,n}^{(i)}\sqrt{v_n}\\
a_{r,n}^{(i)}\sqrt{v_n}
\end{array}
\right) \; ,
\end{equation}
where $v_m=\hbar k_m/m^*$ is the velocity of mode $m$.

Assume now the many impurity scenario of the preceding subsection. As 
hinted in Fig.\ 2b, it is clear that input and output amplitudes of 
successive impurities are related. Indeed, 
\begin{eqnarray}
\label{eqphA}
a_{\ell,m}^{(i)} &=& b_{r,m}^{(i-1)} e^{ik_m(d_i-d_0)}\;, \\
\label{eqphB}
a_{r,m}^{(i)} &=& b_{\ell,m}^{(i+1)} e^{ik_m(d_{i+1}-d_0)} \; . 
\end{eqnarray}
Notice that the distance between impurities is not a constant, $d_i=z_i-z_{i-1}$, 
and that 
the phases in Eqs.\ (\ref{eqphA}) and (\ref{eqphB}) are due to the different 
definition of the asymptotic boundaries for each impurity. 

Repeatedly using Eqs.\ (\ref{scatmat}), (\ref{eqphA}) and (\ref{eqphB})
for the successive impurities we would obtain a linear system of equations relating 
input and output amplitudes of the global disordered region, i.e., between 
first (leftmost) and  ${\cal N}$-th (rightmost) impurities.
This total transmission corresponding to, 
say, unit-amplitude incidence from the left in transverse mode $m_{\rm in}$
to the right in mode $m_{\rm out}$ is simply
\begin{equation}
\label{ttot}
t_{m_{\rm out}m_{\rm in}} \equiv b^{({\cal N})}_{r,m_{\rm out}}
\,\sqrt{
\frac{v_{m_{\rm out}}}
{v_{m_{\rm in}}}
}\; .
\end{equation}
The linear system yielding the output $b$-amplitudes reads 
\onecolumn
\begin{equation}
\label{eq1p5}
\left\{
\begin{array}{rcll}
b_{\ell,m}^{(1)} &=& 
r_{m{m_{\rm in}}}^{(1)} 
+   
\displaystyle\sum_{n}{{t'}_{mn}^{(1)}\, e^{ik_n(d_2-d_0)}\, b_{\ell,n}^{(2)}} 
& (m=1,\dots,M) \\
\rule{0cm}{0.65cm}
b_{r,m}^{(1)} &=& 
t_{m{m_{\rm in}}}^{(1)} 
+   
\displaystyle\sum_{n}{{r'}_{mn}^{(1)}\, e^{ik_n(d_2-d_0)}\, b_{\ell,n}^{(2)}} 
& (m=1,\dots,M) \\
\rule{0cm}{0.65cm}
b_{\ell,m}^{(i)} &=& 
\displaystyle\sum_{n}{r_{mn}^{(i)}\, e^{ik_n(d_i-d_0)}\, b_{r,n}^{(i-1)}}
+   
\displaystyle\sum_{n}{{t'}_{mn}^{(i)}\, e^{ik_n(d_{i+1}-d_0)}\, b_{\ell,n}^{(i+1)}} 
& (i=2,\dots,{\cal N}-1; m=1,\dots,M) \\
\rule{0cm}{0.65cm}
b_{r,m}^{(i)} &=& 
\displaystyle\sum_{n}{t_{mn}^{(i)}\, e^{ik_n(d_i-d_0)}\, b_{r,n}^{(i-1)}}
+   
\displaystyle\sum_{n}{{r'}_{mn}^{(i)}\, e^{ik_n(d_{i+1}-d_0)}\, b_{\ell,n}^{(i+1)}}
& (i=2,\dots,{\cal N}-1; m=1,\dots,M) \\
\rule{0cm}{0.65cm}
b_{\ell,m}^{({\cal N})} &=& 
\displaystyle\sum_{n}{{r}_{mn}^{({\cal N})}\, e^{ik_n(d_{\cal N}-d_0)}\, b_{r,n}^{({\cal N}-1)}}
& (m=1,\dots,M) \\
\rule{0cm}{0.65cm}
b_{r,m}^{({\cal N})} &=& 
\displaystyle\sum_{n}{{t}_{mn}^{({\cal N})}\, e^{ik_n(d_{\cal N}-d_0)}\, b_{r,n}^{({\cal N}-1)}}
& (m=1,\dots,M)
\end{array}\right.\; .
\end{equation}
\twocolumn
Equation (\ref{eq1p5}) is a highly sparse linear system of $2M{\cal N}$ equations.
It can be very efficiently solved with sparse linear solvers like
\verb+ME48+ of the Harwell library \cite{harwell}.

As mentioned in the introduction, a major simplifying property derived from the 
tube symmetry is the transformation relating any two impurities. In fact,
defining a reference impurity at $(z_0,\theta_0)$ the transmission coefficient for 
the $i$-th impurity, positioned at $(z_i,\theta_i)$, reads
\begin{equation}
\label{tit0}
t_{mn}^{(i)} = e^{-i(\theta_i-\theta_0)(\lambda_m - \lambda_n)}  t_{mn}^{(0)}\; ,
\end{equation}
where $\lambda_m$ is the angular momentum of mode $m$. 
The phase in Eq.\ (\ref{tit0}) is due to the rotation needed to link the two impurity
positions. The longitudinal translation does not introduce any additional phase here 
because our definition of the scattering matrix implies a translation of 
the reference boundaries for each impurity. This type of phase does appear
however in Eqs.\ (\ref{eqphA}) and (\ref{eqphB}). 
It can be easily shown that an identical phase 
to that in Eq.\ (\ref{tit0}) is
relating
$r_{mn}^{(i)}$, ${t'}_{mn}^{(i)}$ and ${r'}_{mn}^{(i)}$ with
$r_{mn}^{(0)}$, ${t'}_{mn}^{(0)}$ and ${r'}_{mn}^{(0)}$, respectively.
The overall result is that 
only the reference impurity scattering coefficients are necessary to set up
the linear system Eq.\ (\ref{eq1p5}).

For a given disorder realization we obtain the linear conductance $\tilde{G}$ 
from Eqs.\ (\ref{eq1p5}) and (\ref{ttot}) using Landauer formula to relate conductance
with transmission
\begin{equation}
\label{land}
\tilde{G} = G_0 \sum_{mn}{|t_{mn}|^2}\; .
\end{equation}
As well known, Eq.\ (\ref{land}) corresponds to a two terminal 
measurement and it includes the {\em contact resistance}, i.e., in absence of 
any disorder it yields $G=MG_0$.

\section{Results}

\begin{figure}[t]
\centerline{\includegraphics[width=8.5cm,clip]{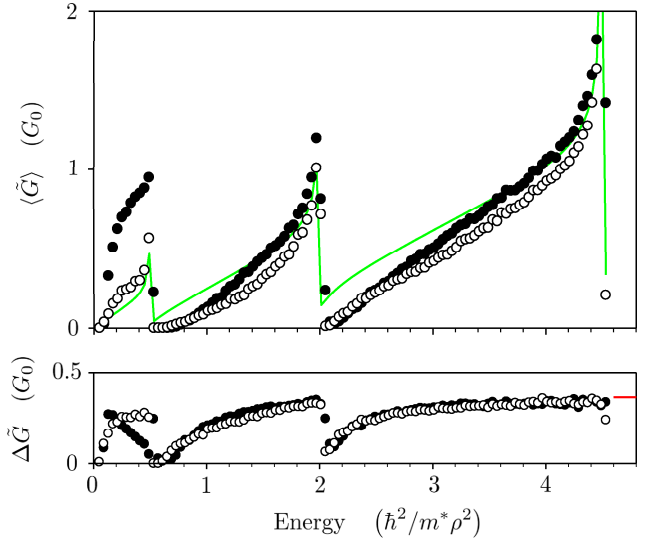}}
\caption{
Mean conductance (upper) and variance (lower) for ${\cal N}=1000$ impurities.
Solid and open symbols correspond to $d_1=0.5\rho$ and $2.5\rho$, respectively.
Other parameters we have used are $d_0=0.5\rho$ and the impurity potential 
given by $\sigma=0.05\rho$ and $V_0=80\; \hbar^2/(m^*\rho^2)$.
The solid line in the upper plot corresponds to the semiclassical 
model and the dash on the right vertical axis of the lower plot gives the universal 
conductance fluctuation value \cite{rev1}.
}
\label{fig3_p}
\end{figure}

\begin{figure}[t]
\centerline{\includegraphics[width=8.50cm,clip]{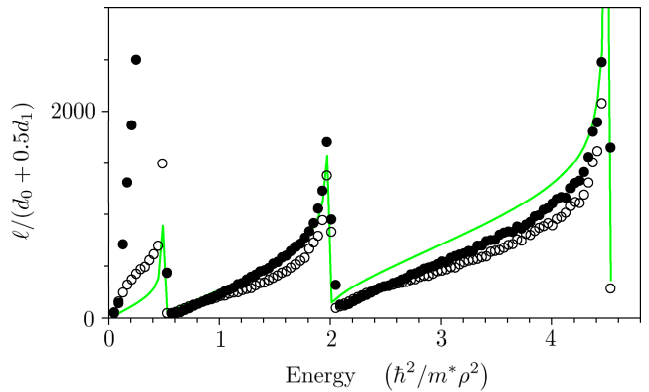}}
\caption{
Localization length, scaled by the mean impurity-impurity distance, for the same parameters
of Fig.\ 3. The solid line shows the corresponding semiclassical result for $\alpha^{-1}$.}
\label{fig4_p}
\end{figure}

Figure \ref{fig3_p} displays the results obtained for a tube having a fixed number
of impurities ${\cal N}=1000$ and an energy range including up to 5 propagating modes.
We have fixed $d_0=0.5\rho$ and the different symbols are for impurity 
distributions characterized by $d_1=0.5\rho$ (solid)
and $d_1=2.5\rho$ (open). The mean conductance has a
conspicuous sawtooth wave behavior, steadily 
increasing with energy until a new mode becomes
propagating and a sharp drop occurs. The variance, $\Delta\tilde{G}\equiv(\langle\tilde{G}^2\rangle
-\langle\tilde{G}\rangle^2)^{1/2}$, shows similar drops 
and a clear convergence with increasing energy 
towards the universal conductance fluctuation value. 
Except for the first {\em plateau} \cite{plateau}
the results for $d_1=0.5\rho$ and $2.5\rho$ look quite similar. 

As mentioned in the introduction, the key to analyze the 
results of Fig.\ \ref{fig3_p} is the localization length $\ell$. 
Figure \ref{fig4_p} displays this variable for the same parameters of Fig.\ \ref{fig3_p}.
We have found $\ell$ by disorder averaging the Lyapunov exponent in a long wire. More specifically, 
we have done the calculations for  $10^4$ impurities, for which the wire length is much
larger than $\ell$ for all energies shown in Fig.\ \ref{fig4_p}.
There is a qualitative similarity between Figs.\ \ref{fig3_p} and
\ref{fig4_p}. The energy dependence of the conductance for a fixed wire
length can be interpreted in terms of the strong variation of the 
localization length with the energy. In general, as the energy increases 
the wire evolves from the localized towards the metallic regime until 
the appearance of a new mode causes a sudden drop back towards the  
localized case.

The first plateau, with only one propagating mode, shows a qualitative difference 
from the others. Remarkably, the mean conductance with $1000$ impurities is close 
to the maximal value $G_0$ when $d_1=0.5\rho$. For $d_1=2.5\rho$ it also shows a 
different energy dependence with a faster energy increase than for other plateaus. 
This behavior can be
attributed to the effective one dimensionality in this limit. Indeed, it can be shown 
that the model is equivalent to a Kronig-Penney lattice in which the spacing between 
1D barriers is not exactly regular but has some fluctuation around the mean positions.
For a fixed $d_0$ regularity is inversely proportional to $d_1$. The more regular
the distribution, the higher the conductance due to the formation of 
extended Bloch states. This is nicely confirmed by the localization length (Fig.\ \ref{fig4_p}) 
which shows a dramatic enhancement in the first plateau for the smaller $d_1$.
However, for large enough wire lengths the states are always localized, i.e., 
$\ell$ is large but remains finite. 

It is also worth mentioning that the localization
length scales, approximately, with the mean impurity-impurity distance. As Fig.\ \ref{fig4_p}
shows, this is clear at the beginning of the second and third plateaus, where 
a linear increase of the localization length with energy is found.
This linear behavior agrees with a prediction discussed by Thouless in Ref.\ \cite{thouless}.
In this reference the relation between localization length and density of states is 
stressed, showing that for a particular density of states,
a {\em Gaussian white noise} model \cite{halperin}, one finds $\ell\sim E$. 
The discontinuities in localization
length (Fig.\ \ref{fig4_p}) could then indicate the existence of discontinuities in the 
density of states of the disordered system. Such discontinuities would not be surprising
since the density of states of the clean quasi-1D system diverges
as $(E-\varepsilon_m)^{-1/2}$ at the beginning of each plateau, when $E\approx\varepsilon_m$.

\begin{figure}[t]
\centerline{\includegraphics[width=8.0cm,clip]{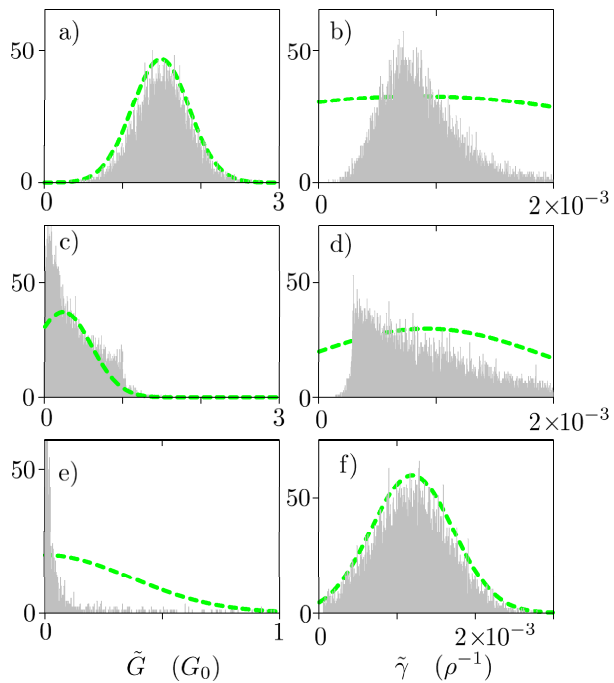}}
\caption{Probability distributions of the conductance (left panels) and Lyapunov exponents (right).
Ordered by rows the results correspond to increasing wire
lengths: 200 impurities in (a) and (b), 1000 in (c) and (d) and 5000 in
(e) and (f). The dashed lines show the normal distributions,
Eqs.\ (\ref{pd1}) and (\ref{pd2}), from the theory of disordered conductors
using an arbitrary vertical scale.
As expected, Gaussian distributions are found
in the conductance of the metallic regime (a), and in the Lyapunov exponent 
of the localized regime (f).}
\label{fig5_p}
\end{figure}

In Fig.\ \ref{fig5_p} we show the probability distributions of the conductance $\tilde{G}$ and 
Lyapunov exponent $\tilde\gamma$ for a fixed energy $E=1.6\hbar^2/{(m^*\rho^2)}$
and three different wire lengths, corresponding to 200, 1000 and 5000 impurities and
$d_1=2.5\rho$.
Comparison with the prediction of the theory of disordered conductors, 
Eqs.\ (\ref{pd1}) and (\ref{pd2}),
shows the deviation in the intermediate regime, ${\cal N}=1000$, when $L\approx 1.5\ell$
and the wire is neither in localized nor in metallic regime. In the limiting cases, 
however, either the conductance (metallic) or the Lyapunov exponent (localized) closely 
follows the expected behavior.

Figures \ref{fig3_p} and \ref{fig4_p} also show the semiclassical result
in each case (continuous line). As mentioned in the Introduction, the 
semiclassical approach neglects
interference effects by directly composing the transmission probabilities, as opposed
to the phase-coherent composition of transmission amplitudes 
in fully quantum mechanical calculations.
Notice that if all phases of Eqs.\ (\ref{eqphA}), (\ref{eqphB}), (\ref{eq1p5})
and (\ref{tit0}) are neglected the formalism does not depend
on the impurity positions $(z_i,\theta_i)$, only on its total number ${\cal N}$.
This causes Ohmic behavior, characterized by parameter $\alpha$ in Eq.\ (\ref{Gsc}).
It is also possible to relate $\alpha$ with the transmission probability of a single 
impurity $T^{(1)}$. Indeed, using the composition rule of incoherent 
scatterers we find the total transmission $T$ from 
\begin{equation}
\frac{1}{T}-\frac{1}{M} = {\cal N}\left( \frac{1}{T^{(1)}}-\frac{1}{M}  \right)\; ,
\end{equation}
leading to the explicit expression
\begin{equation}
\label{scd}
\alpha = \left( \frac{1}{T^{(1)}}-\frac{1}{M}  \right)\frac{1}{d_0+0.5d_1}\; .
\end{equation}

As seen in Fig.\ \ref{fig4_p}, $\alpha^{-1}$ is close to the 
localization length or, equivalently, $\alpha$ is close to the Lyapunov exponent.
However, both values do not coincide in general, making the comparison between 
quantum and classical conductance, Eq.\ (\ref{diff}), rather involved. Another
information we can extract from the semiclassical result regards the discontinuities
in localization length as a function of energy. Besides Thouless' argument on the 
density of states given above, we also expect similar drops in the semiclassical 
length $\alpha^{-1}$ from Eq.\ (\ref{scd}). Indeed, the dependence of $T^{(1)}$ on energy
is a staircase with rounded edges. This smooth energy dependence is broken by the explicit
appearance of $M$, the number of modes, in Eq.\ (\ref{scd}). A sharp increase in $\alpha$
occurs when the energy allows a new mode but $T^{(1)}$ has not yet increased.

\section{Conclusions}

We have studied the quantum transport properties of tubular shaped nanowires with 
randomly distributed impurities. 
For a given wire length our calculations predict a sawtooth-like
behavior of the conductance as a function of energy; conspicuous drops appearing
when the energy is just barely enough to allow the propagation of a new transverse
mode. The analysis of our results within the scaling theory of disordered conductors
explains the observed behavior as a general tendency of the wire to evolve from localized
towards metallic regimes with increasing energy. Abrupt changes in the density of states
occur when a new mode becomes propagating, leading the wire back towards a more localized regime.

We have also calculated the localization length of the disordered wire for energies
covering from one to five propagating modes. Qualitative interpretation in terms of the 
density of states explains the linear-with-energy dependence at the beginning 
of the conducting plateaus, as well as the discontinuous drops already mentioned. 
The semiclassical description has been also discussed and shown to
provide a complementary interpretation of the discontinuous drops. In the metallic 
regime, the comparison between the average quantum conductance and the semiclassical 
one is complicated when the Lyapunov exponent and the semiclassical coefficient for Ohmic 
behavior do not coincide.

\section*{Acknowledgments}
L.S.\ was supported by the MEC (Spain) within its program for sabbaticals and 
Grant (FIS2008-00781).
M.-S.C.\ was supported by the KOSEF Grant (2009-0080453).

\appendix

\section{Single impurity scattering}
The formalism of Sec.\ \ref{scatsec} requires the scattering 
matrix of a single impurity as an input. We find this by solving 
the Schr\"odinger equation for the corresponding open boundary 
problem
using the quantum-transmitting boundary 
algorithm \cite{qtbm}. The wave function
fulfills the equation
\begin{eqnarray}
\label{eqsch}
\left(
-\frac{\hbar^2}{2m^*}\frac{d^2}{dz^2}
-\frac{\hbar^2}{2m^*\rho^2}\frac{d^2}{d\theta^2}
+ V(z,\theta)
\right)
&\psi(z,\theta)& = \nonumber\\ 
E&\psi(z,\theta)&\; ,
\end{eqnarray}
where $m^*$ is the electron's effective mass and $V(z,\theta)$ is given 
by Eq.\ (\ref{vimpu}) with the impurity at $(z_i,\theta_i)=(0,0)$.
Far from the impurity, in the parts of the tube that play the role of contact
leads (or contacts for short),
the potential vanishes and the Hamiltonian 
separates in longitudinal and angular (transverse) contributions. The 
angular problem
\begin{equation}
-\frac{\hbar^2}{2m^*\rho^2}\frac{d^2\phi_n(\theta)}{d\theta^2}
 =
\varepsilon_n \phi_n(\theta)\; ,
\end{equation}
yields the set of transverse modes $\{\phi_n,\varepsilon_n\}$,
\begin{eqnarray}
\label{basisf}
\phi_n(\theta) &=& \frac{1}{\sqrt{2\pi}}e^{i\lambda_n\theta}\quad (\lambda_n=0,\pm1,\dots)\; , \\
\varepsilon_n &=&
\label{basise}
\frac{\hbar^2\lambda_n^2}{2m^*\rho^2} \;.
\end{eqnarray}

Equation (\ref{eqsch}) can be solved using finite differences in a Cartesian grid. However, 
increasing the number of grid points becomes very costly and this method lacks accuracy 
in the angular integrations with oscillating functions like those of Eq.\ (\ref{basisf}).
Therefore, we have used a mixed approach, in which the $z$ coordinate is discretized
in a uniform grid
while the angular dependence is described by expanding in the angular eigenfunctions,
i.e.,
\begin{equation}
\label{angexp}
\psi(z,\theta) = \sum_n{\psi_n(z)\,\phi_n(\theta)}\; .
\end{equation}
The unknown band amplitudes $\psi_n(z)$ fulfill coupled-channel equations
\begin{equation}
\label{eccm0}
-\frac{\hbar^2}{2m^*}
\psi_{n}^{''}(z)
+(\varepsilon_{n}-E)\,
\psi_{n}(z)
=
-\sum_{n'}{V_{nn'}(z)\psi_{n'}(z)}\;,
\end{equation}
where
\begin{equation}
V_{nn'}(z)=
\frac{1}{2\pi}\int{d\theta\, V(z,\theta)e^{i(\lambda_{n}-\lambda_{n'})\theta}}
\; .
\end{equation}

In the contacts the band amplitudes take a similar form to Eq.\ (\ref{wf}),
\begin{equation}
\label{eccm1}
\psi_n(z) =
a_{c,n} e^{i s_c k_n (z-z_c)}
+
b_{c,n} e^{-i s_c k_n (z-z_c)}
\;.
\end{equation}
This expression is for a propagating channel, for which $\varepsilon_n<E$ and 
$k_n=\sqrt{2m^*(E-\varepsilon_n)}/\hbar$ is a real number.
It also applies to evanescent ones, $\varepsilon_n>E$,
if we assume in this case $a_{c,n}=0$ and a purely imaginary wavenumber 
$k_n=i\sqrt{2m^*(\varepsilon_n-E)}/\hbar$.
Notice that the output amplitudes can be obtained from the wave function right at 
the contact position,
\begin{equation}
\label{eqbcn}
b_{c,n} = \psi_n(z_c) - a_{c,n}\; .
\end{equation}
Substituting Eq.\ (\ref{eqbcn}) in Eq.\ (\ref{eccm1}) we obtain 
\begin{eqnarray}
\label{eccm2}
\psi_n(z)-\psi_n(z_c)\,e^{-i s_c k_n (z-z_c)}
=&&\nonumber\\
2 i a_{c,n}  \sin&&\!\!\!(s_c k_n(z-z_c))\,,
\end{eqnarray}
which is the quantum-transmitting-boundary equation for the contacts.

Equations (\ref{eccm0}) and (\ref{eccm2}), for the central and contact regions,
respectively, form a closed set which does not invoke the wave function
at any external point. Of course, this is not true for any of these two subsets separately, 
since central and contact regions are connected 
through the derivative in Eq.\ (\ref{eccm0}) and of $\psi_n(z_c)$ in Eq.\ (\ref{eccm2}).
In practice, we use a uniform grid in $z$ with $n$-point formulae for the derivatives ($n\approx5-11$)
and truncate the expansion in transverse bands, Eq.\ (\ref{angexp}), to include typically 50-100 terms. 
The resulting sparse linear problem is then solved as in Sec.\ \ref{scatsec} using 
routine \verb+ME48+ \cite{harwell}.


\begin{thebibliography}{}
\bibitem{Anderson} P.W. Anderson, Phys.\ Rev.\ {\bf 109}, 1492 (1958)
\bibitem{scaling} P.W.\ Anderson, D.J.\ Thouless, E. Abrahams, D.S.\ Fisher, 
Phys. Rev. B {\bf 22}, 3519 (1980)
\bibitem{rev1} C.W.J.\ Beenakker, Rev.\ Mod.\ Phys.\ {\bf 69}, 731 (1997)
\bibitem{rev2} L.I.\ Deych, A.A.\ Lisyansky, and B.L.\ Altshuler,
Phys. Rev. B {\bf 64}, 224202 (2001)
\bibitem{rev3} J.B.\ Pendry, Advances in Physics {\bf 43}, 461 (1994)
\bibitem{rev4} B. Kramer and A. MacKinnon, Rep.\ Prog.\ Phys.\ {\bf 56}, 1469 (1993).
 \bibitem{expertube}
V.Y.\ Prinz, V.A.\ Seleznev, A.K.\ Gutakovsky, A.V.\ Chehovskiy,
V.V.\ Preobrazhenskii, M.A.\ Putyato, and T.A.\ Gavrilova, 
Physica E {\bf 6}, 828 (2000)
\bibitem{et2}
O.G. Schmidt and K. Eberl, Nature (London) {\bf 410}, 168 (2001)
\bibitem{et22}
A. Lorke, S. B\"ohm, and W. Wegscheider, Superlattices Microstruct.\
{\bf 33}, 347 (2003)
\bibitem{et3}
N. Shaji, H. Qin, R.H.\ Blick, L.J.\ Klein, C. Deneke, and O.G.\
Schmidt, Appl.\ Phys.\ Lett.\ {\bf 90}, 042101 (2007)
\bibitem{et4} M. Jung, J.S.\ Lee, W. Song, Y.H.\ Kim,
S.D.\ Lee, N. Kim, J. Park, M.-S.\ Choi,
S. Katsumoto, H. Lee, J. Kim,
Nanoletters {\bf 8}, 3189 (2008)
\bibitem{theotube} G. Ferrari, A. Bertoni, G. Goldoni, and E. Molinari
Phys.\ Rev.\ B {\bf 78}, 115326 (2008)
\bibitem{tt2} M. Trushin and J. Schliemann,
New Journal of Physics {\bf 9}, 346 (2007)
\bibitem{tt3}L.I.\ Magarill, D.A.\ Romanov, A.V.\ Chaplik,
JETP {\bf 86}, 771 (1998)
\bibitem{cahay} M. Cahay, M. McLennan, S. Datta,
Phys.\ Rev.\ B {\bf 37}, 10125 (1988)
\bibitem{notesc} This result, discussed at the end of Sec.\ 3, is exactly 
fulfilled in the tube symmetry. It is not evident in other geometries like
the planar one (see Ref.\ \cite{cahay})
\bibitem{thouless} D.J.\ Thouless, J. Phys.\ C {\bf 5}, 77 (1972)
\bibitem{datta} S. Datta, {\em Electronic transport in mesoscopic systems}
(Cambridge University Press, 1997)
\bibitem{harwell} HSL, A Collection of Fortran codes for large-scale scientific 
computation. See http://www.hsl.rl.ac.uk, (2007)
\bibitem{plateau} Although the energy dependence of the conductance with disorder 
no longer resembles a staircase 
we still refer by plateaus to the different energy intervals in which the number of modes
is constant. For instance, the first plateau, with just one mode, corresponds to
$0< E< 0.5\,\hbar^2/(m^*\rho^2)$.
\bibitem{halperin} B.I.\ Halperin, Phys.\ Rev.\ {\bf 139}, A104 (1965)
\bibitem{qtbm} C.S.\ Lent, D.J. Kirkner, J. Appl.\ Phys.\ {\bf 67}, 6353 (1990)
\end{thebibliography}
\end{document}